# Successive Phase Transitions from a Composite Fermion Liquid to a Fractional Quantum Hall State at $\nu = 3/2$ Driven by In-Plane Magnetic Field

Xinghao Wang[1], L. N. Pfeiffer[2], A. Gupta[2], K. W. Baldwin[2], K. W. West[2], and Rui-Rui Du[1*]

[1]International Center for Quantum Materials, School of Physics, Peking University; Beijing 100871, China

[2]Department of Electrical Engineering, Princeton University, Princeton, NJ 08544, USA

[*]Corresponding author. Email: rrd@pku.edu.cn

**ABSTRACT**

**We report an even-denominator fractional quantum Hall state at $\nu = 3/2$ induced entirely by in-plane magnetic field $B_{\parallel}$ in an ultra-high-mobility GaAs quantum well. As $B_{\parallel}$ increases, the system undergoes two successive transitions: from a composite fermion liquid to a soft-gap FQH state ($B_{\parallel} \sim 12.2\ T$), then via a topological phase transition to a hard-gap robust FQH state ($B_{\parallel} \sim 14.7\ T$), accompanied by a daughter state at $\nu = 19/13$. We present systematic data, and discuss a possible scenario in interpreting these findings. Our work demonstrates that topological order may be engineered through $k$-space Fermi contour splitting under an in-plane magnetic field.**

***Introduction*** – Even-denominator fractional quantum Hall (FQH) states in GaAs/AlGaAs quantum wells (QWs) at half-integer filling factors have attracted considerable attention because their emergent quasiparticles are expected to exhibit non-Abelian statistics [1-3], making them promising candidates for topological quantum computation [4]. Conventionally, such states appear in excited Landau levels (LLs) - most notably the half-filled $N = 1$ second Landau level (SLL) - where the nodal structure of the electronic wavefunction weakens the short-range Coulomb repulsion and thereby enables pairing among composite fermions (CFs) [5,6]. Such mechanism is universal in many other materials [7-13]. By contrast, in the $N = 0$ lowest Landau level (LLL), the half-filled states at $\nu = 1/2$ and $3/2$ are typically compressible, characterized by a CF Fermi sea, while incompressible fractional states emerge only at nearby odd-denominator fillings [6].

Nevertheless, an exception was discovered in 1992 when a robust $\nu = 1/2$ FQH state was observed in a high-mobility two-dimensional electron gas (2DEG) confined to either wide single QWs [14] or closely coupled double QWs [15], and a $\nu = 3/2$ FQH state was found as well [16]. In the double-well case - where inter-well tunneling is negligible - the $\nu = 1/2$ FQH state is well understood as the two-component Abelian Halperin-Laughlin $\Psi_{331}$ state [16-18]. The nature of the half-filled state in wide single quantum wells, however, has remained somewhat perplexing. Early proposals suggested a one-component Pfaffian wavefunction [19,20], but subsequent work argued instead for a two-component description [16,17,21-23]. More recently, a growing body of evidence has swung the consensus back toward a single-component non-Abelian Pfaffian ground state [24-26].

The crucial development in understanding the nature of the $\nu = 1/2$ state comes from examining its hierarchical descendants. According to Levin and Halperin's theoretical framework, the Pfaffian parent state supports a sequence of daughter states at specific fractional fillings, with the simplest ones occurring at $\nu = 8/17$ and $7/13$ [26-30]. These daughter states are topologically distinct from the conventional Jain-sequence composite-fermion states that appear at the same filling factors [27,30,31]. Notably, the daughter states offer a hierarchical pathway through which the Pfaffian character of the $\nu = 1/2$ ground state can be probed indirectly. Theoretical analysis showed that all daughters of paired states at $\nu = 1/2$, e. g., the daughters of the anti-Pfaffian state are $\nu = 6/13$ and $9/17$, and found that no two distinct parent states share an identical set of daughter states [28]. The unique parentage implies that electrical transport measurements alone could determine the topological order of even-denominator FQH states.

Two experimental knobs are available to tune the $\nu = 1/2$ state in wide quantum wells. The first is electron density tuning. By adjusting the front and back gate voltages, one can vary the carrier density $n$ symmetrically, which in turn reshapes the charge distribution and modulates the symmetric-antisymmetric subband splitting $\Delta_{SAS}$ [14,32]. Upon increasing $n$, the charge distribution becomes increasingly bilayer-like, $\Delta_{SAS}$ decreases, and an intermediate-density window opens where the $\nu = 1/2$ state together with its daughter states becomes incompressible [27,30]. The second knob is in-plane magnetic field. Tilting the sample relative to the total field introduces a parallel component $B_{\parallel}$ [33], which couples to the out-of-plane orbital motion and suppresses interlayer tunneling, thereby also reducing $\Delta_{SAS}$ [31,34]. At a fixed low density where the surrounding fractions follow the Jain sequence at zero tilt, progressively larger $B_{\parallel}$ induces the same sequence of transitions: the $1/2$ state strengthens, and its daughter states suddenly emerge with enhanced robustness [31]. Another effect of in-plane magnetic field applied to a finite-width quantum well is to destabilize uniform pairing and promote anisotropic phases such as stripe or nematic phases due to the breaking of spatial rotational symmetry in the Coulomb interaction [35-38].

In this manuscript, we report an even-denominator FQH state at $\nu = 3/2$ completely driven by a strong in-plane magnetic field. The sample has low density and a medium-width QW, which distinguishes it from the wide QW samples discussed above since only the symmetric subband is occupied and the system remains single-layer at zero tilt and even at relatively large angles $\theta \sim 70°$. Instead of weakening $\Delta_{SAS}$ and suppressing interlayer tunneling, the strong in-plane magnetic field splits the Fermi contour into two disconnected parts, signaling the field-induced one-component to two-component transition [39,40]. As the Fermi surface splits into two disconnected droplets in $k$-space, the real-space electron distribution can be viewed as a bilayer due to a $k_y - z$ locking (if the in-plane field is along $x$ direction), although it retains significant central overlap because the out-of-plane magnetic length ($l_{B_{||}} \sim 6nm$) is comparable to the wavefunction separation. This effect might induce Abelian or non-Abelian states at even-denominator filling factors. Additionally, a single daughter state at $\nu = 19/13$ is found, which could help pinpoint the topological order of the $\nu = 3/2$ state.

***Fractional quantum Hall state at*** $\boldsymbol{\nu = 3/2}$ – The sample we use is a van de Pauw device of a GaAs/AlGaAs quantum well with a width of $w = 40.5nm$. The 2DEG has a low density of $n = 1.4 \times 10^{11} cm^{-2}$ and an ultra-high mobility of $\mu = 5.0 \times 10^{7} cm^{2}/Vs$ after brief illumination with a red LED. The contacts are made of In/Sn alloy annealed in a $N_2/H_2$ mixture at 450°C. The sample was measured in a dilution refrigerator with an $18T$ magnetic field and base temperature around $25mK$. The sample was mounted on a rotatable sample holder allowing the tilt angle $\theta$ to be tuned in situ. In-plane magnetic field $B_{||} = B_{tot} sin\theta$ can continuously tune the topological phase while out-of-plane magnetic field $B_{\perp} = B_{tot} cos\theta$ determines the filling factor of LLs.

Fig. 1(a) shows the magnetoresistance and Hall resistance traces without in-plane magnetic field, which are regular traces for an ultra-high-mobility 2DEG in a GaAs QW. FQH state in the SLL at $\nu = 7/2, 8/3, 5/2, 7/3$ indicates the high quality of our sample and the Jain sequence as the IQH state of CFs appears around $\nu = 3/2$.

The phases at filling factor $\nu = 5/2$ and $\nu = 3/2$ are sensitive to in-plane magnetic field. The main role of the in-plane magnetic field at $\nu = 5/2$ is the induced anisotropy of the CF Fermi surface. Since non-Abelian FQH states such as the Moore-Read state and the anti-Pfaffian state are extremely sensitive to anisotropy of Fermi contour, the $\nu = 5/2$ FQH state gradually turns into a CF liquid at $\theta = 76.2°$, similar to the situation of LLL at zero tilt (see Fig. 1(b)).

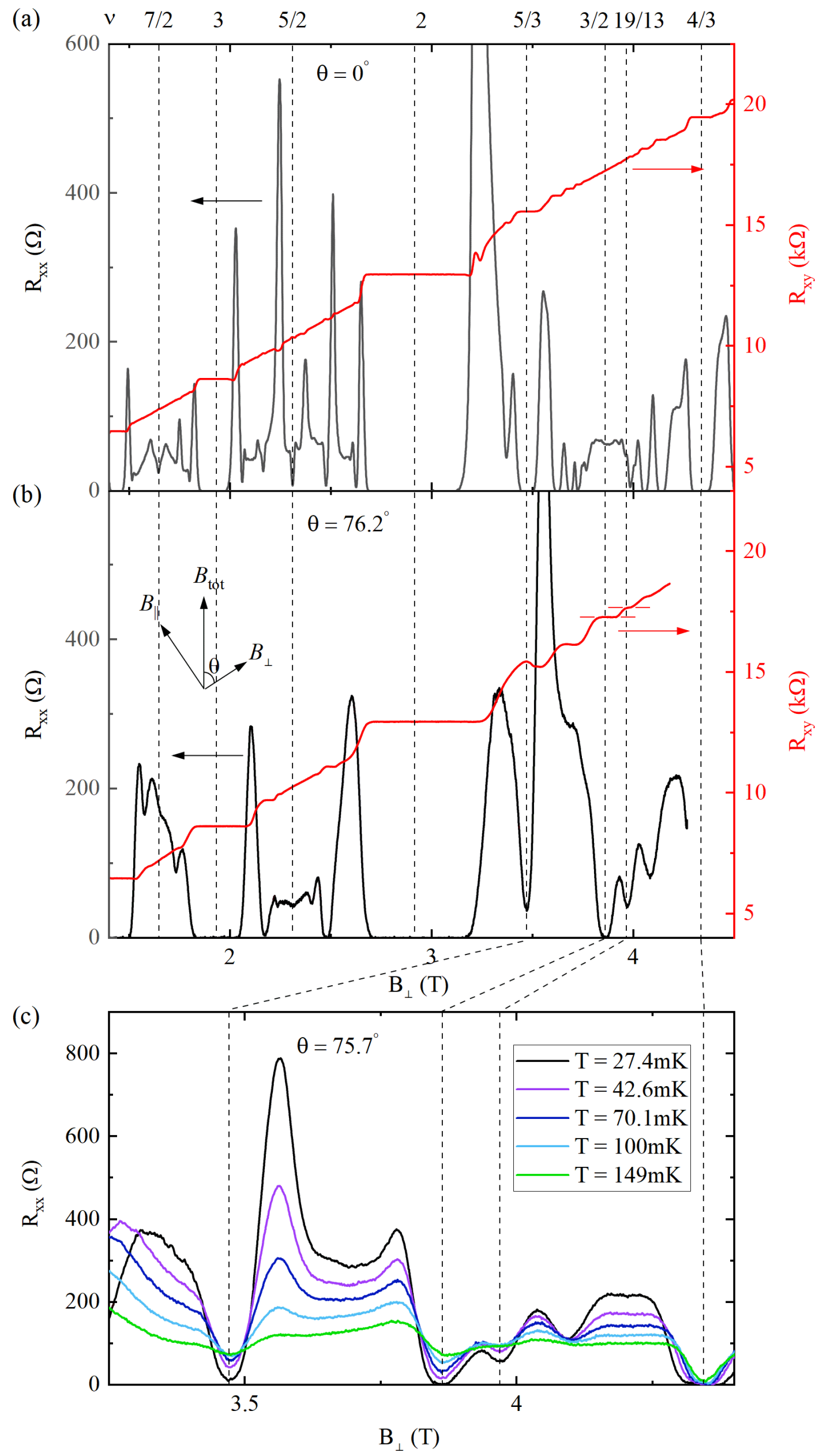

FIG. 1. Magnetoresistance $R_{xx}$ and Hall resistance $R_{xy}$ measured at (a) $\theta = 0°$ and (b) $\theta = 76.2°$ at a base temperature of $T \sim 25mK$. Hall plateaus at $\nu = 3/2$ and $19/13$ are marked.

The inset of panel (b) illustrates the schematic of the tilted field configuration. (c) Magnetoresistance $R_{xx}$ measured at $\theta = 75.7°$ at varying temperatures.

At $\nu = 3/2$, however, a robust FQH state is driven by the tilt magnetic field, with $R_{xx}$ vanishing and a quantized Hall plateau of $R_{xy} = 2h/3e^2$ (see Fig. 1(b)). The FQH state at $\nu = 5/3$ becomes weaker compared with the trace at $\theta = 0°$ and a developing reentrant IQH state near $\nu = 1.6$. Apart from these, an exotic FQH at $\nu = 19/13 = 1 + 6/13$ accompanies the emergence of the $\nu = 3/2$ FQH state. This state is possibly a daughter state for $\nu = 3/2$ FQH state [28], which hints that the parent state at $\nu = 3/2$ could be anti-(331) or anti-Pfaffian depending on different pictures. The former candidate is the particle-hole conjugate of (331) state at $\nu = 1/2$. Topology of $\nu = 3/2$ FQH state cannot be uniquely determined since the daughter state on the higher-fillings side is affected by the surrounding insulating phase. This case is quite usual since the only observed daughter state of $\nu = 5/2$ is $\nu = 2 + 6/13$ [41].

Fig. 1(c) shows how the FQH state at $\nu = 3/2$ and $\nu = 19/13$ evolves with changing temperature at $\theta = 75.7°$. These data demonstrate that both FQH states exhibit a thermal excitation gap, while the next $R_{xx}$ minimum near filling factor $\nu = 1.42$ does not show such behavior probably due to its negligible energy gap. We also see the phase adjacent to $\nu = 5/3$ weakens with increasing temperature.

***Phase transitions at $\nu = 3/2$*** – Fig. 2 shows how magnetoresistance and Hall resistance evolve with tilt angle $\theta$. From $\theta = 0°$ to around $70°$ (black and blue traces), $R_{xx}$ near $\nu = 3/2$ follows the normal pattern of Jain sequence with a CF Fermi sea at $\nu = 3/2$, which is the ground state in the LLL. Anisotropy induced by in-plane magnetic field is the dominant effect.

When the in-plane magnetic field abruptly increases to around $B_{||} = 12T$ and $\theta = 72°$, a developing soft-gap FQH state emerges. The $R_{xx}$ minimum remains finite at base temperature and a nearly quantized plateau of $R_{xy}$ is developing (orange traces at $\theta = 74.4°$). This even-denominator FQH state is accompanied by a normal Jain sequence with robust $\nu = 5/3, 4/3$ FQH states and developing $\nu = 8/5, 7/5$, which is different from the scenario of robust $\nu = 3/2$ FQH state in higher in-plane magnetic field. As the in-plane field goes higher, the robust $\nu = 3/2$ FQH state emerges (red traces at $\theta = 76.2°$). The FQH state $\nu = 5/3$ becomes weaker as in-plane magnetic field increases. It needs to be explored how the in-plane magnetic field drives the phase transition from a CF liquid to a FQH state.

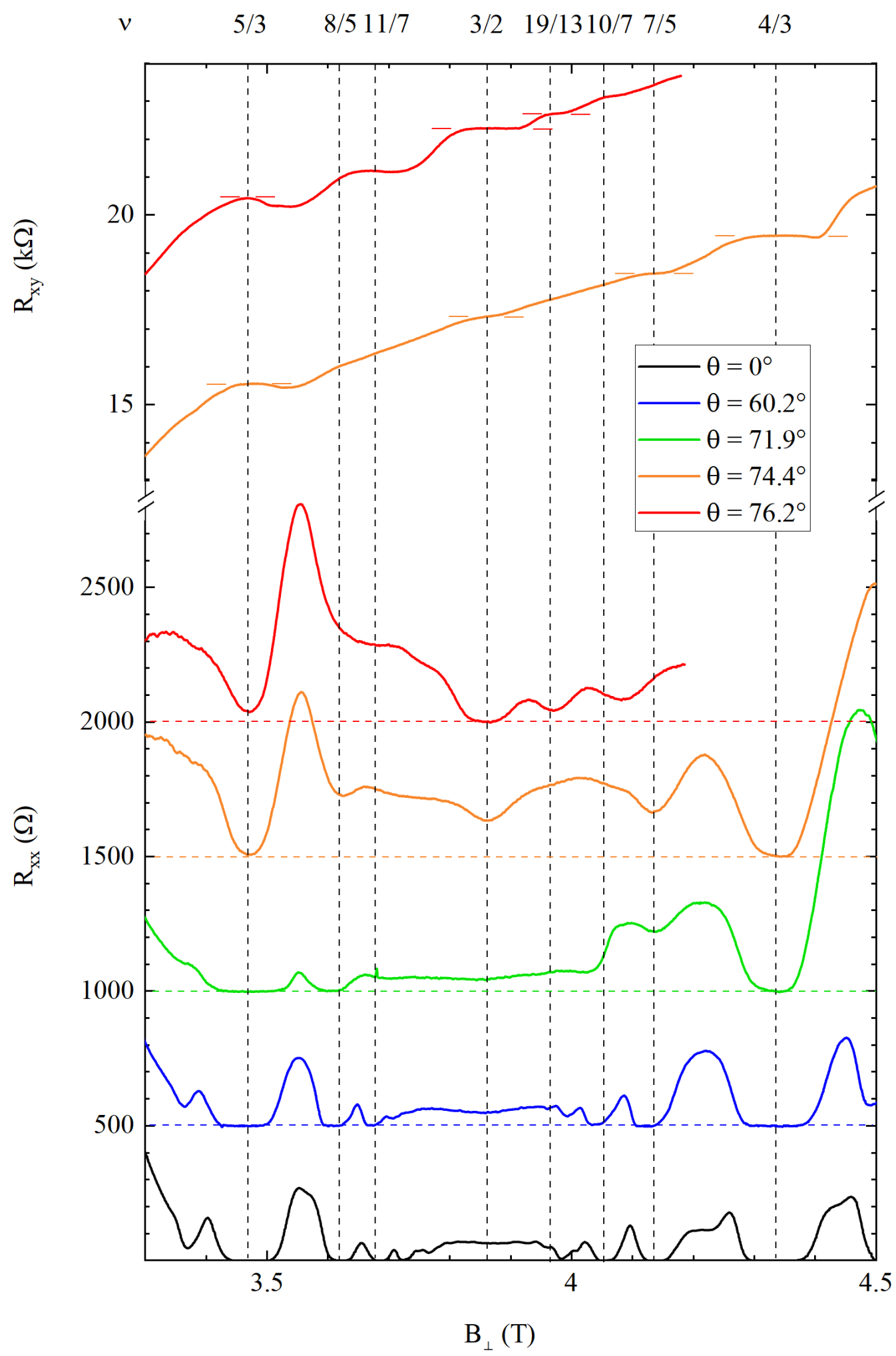

FIG. 2. Magnetoresistance $R_{xx}$ and Hall resistance $R_{xy}$ measured at various tilt angles ranging

from $\theta = 0°$ to $76.2°$ at a base temperature of $T \sim 25mK$. Hall plateaus at $\nu = 5/3$, $3/2, 19/13$ and $4/3$ are indicated. A resistance jump at $B_{\perp} = 4.06T$ in the trace of $\theta = 71.9°$ signals a phase transition induced by the in-plane magnetic field. Two distinct phases at $\nu = 3/2$ are revealed by the traces at $\theta = 74.4°$ and $76.2°$.

The phase at $\nu = 3/2$ has two transition points as in-plane magnetic field increases from zero (Fig. 3(a)). The first transition point at $B_{||} = 12.2T$ marks a phase transition from CF Fermi liquid to a soft-gap developing FQH state. A sudden resistance change near $B_{||} = 12.2T$ coincides with the emergence of the developing $\nu = 3/2$ FQH state (see Fig. 3(b)). A possible picture could explain the emergence of the sudden resistance change. The resistance change may reflect a Fermi-surface transition in which the "peanut"-shaped contour (left panel of Fig. 3(d)) bifurcates into two disconnected parts (middle panel of Fig. 3(d)), arising from a $k_y$-dependent magnetic potential that flips the curvature of the subband dispersion [39] (The $y$ direction denotes the in-plane orientation perpendicular to $B_{||}$). The splitting is accompanied by a logarithmic van Hove singularity in the density of states and a divergent cyclotron effective mass, accounting for the resistance jump at the phase transition point.

For the soft-gap phase at $\nu = 3/2$, the energy scale of the excitation gap of the soft-gap state is determined with the method introduced in Ref. [37] since a developing FQH state often shows a limited range of activation. This method is to measure the ratio of resistance minimum and the average peak (see Fig. 3(e)), $S = 2R_{min}/(R_{p1} + R_{p2})$, and the energy gap $\Delta^S$ is determined from $S \propto exp\,(-\Delta^S/2k_BT)$. For the soft-gap FQH state at $\nu = 3/2$, $\Delta^S \sim 30mK$, which could be quantitatively different from the activation energy gap.

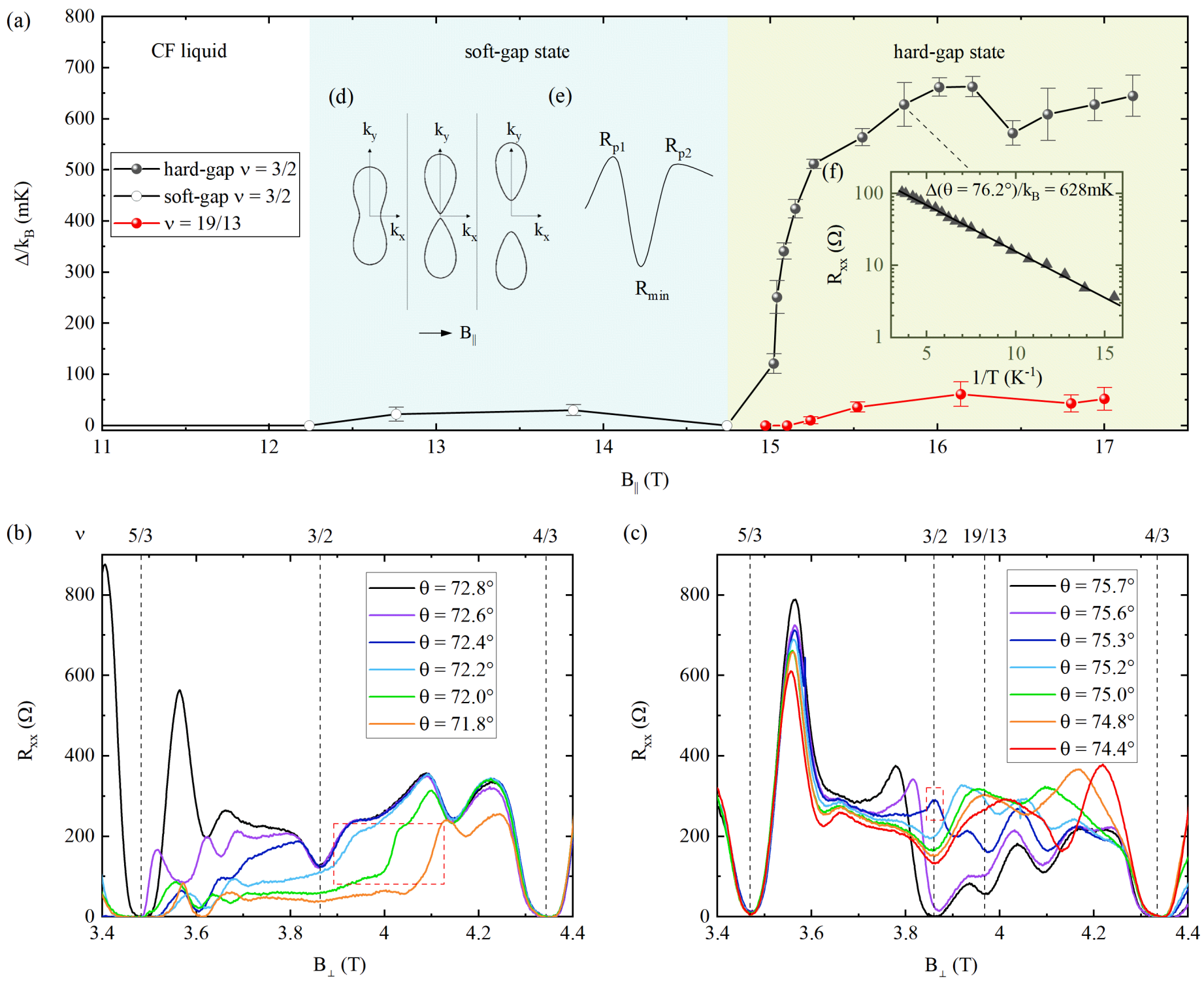

FIG. 3. (a) Thermal excitation gap energy at $\nu = 3/2$ and $19/13$ as a function of the in-plane magnetic field $B_{||}$. Successive phase transitions of the $\nu = 3/2$ state occur at $B_{||} = 12.2T$ and $14.7T$. (b) Magnetoresistance near the phase transition from the CF-liquid to the developing soft-gap FQH state at the base temperature. The red dashed box indicates the resistance jump induced by $B_{||}$. (c) Magnetoresistance near the phase transition from the developing soft-gap FQH state to the hard-gap FQH state at the base temperature. The red dashed box marks the gap closure. Inset (d) illustrates the Fermi contour transition driven by $B_{||}$. Inset (e) shows a schematic of the energy gap measurement using the valley-peak ratio. Inset (f) presents the Arrhenius plot at $\theta = 76.2°$.

The second phase transition at $B_{||} = 14.7T$ marks a phase transition from a soft-gap developing FQH state to a hard-gap robust FQH state. The activation energy gap of the hard-gap FQH state is measured through an Arrhenius plot (see Fig. 3(f)). The gap increases sharply with in-plane magnetic field and saturates when $B_{||} > 15.5T$. It should be noticed that $\nu = 19/13$ daughter state also emerges with the hard-gap $\nu = 3/2$ FQH state. To examine the phase transition in detail, magnetoresistance from $\theta =$ $74.4°$ to $75.7°$ (Fig. 3(c)) demonstrates that the resistance minimum first rises from nearly zero to $\sim 300\Omega$ and then falls to $\sim 130\Omega$, and the energy gap at $\nu = 3/2$ closes at $\theta = 75.3°$, which is equivalent to $B_{||} = 14.7T$. The gap closure suggests a topological phase transition from a soft-gap FQH state to a hard-gap FQH state. The resistance dip of $\nu = 19/13$ still survives at $\theta = 75.3°$ because the in-plane field at this filling factor is relatively higher, which equals $B_{||} = 15.1T$.

***Discussion*** – The topological character of both the soft-gap and hard-gap $\nu = 3/2$ FQH states remains an open and fascinating question, not least because the very appearance of even-denominator FQH states in this regime, together with the associated phase transitions, came as a surprise. In the following we outline one possible scenario built entirely on the Fermi-surface restructuring driven by the in-plane magnetic field; this picture is offered as a tentative guide rather than a definitive account, and its validity awaits detailed theoretical and numerical investigation. Further theoretical and computational work will be needed to judge whether this framework captures the essential physics of the experiment or whether alternative interpretations prove more appropriate.

The topological phase transition between the soft-gap and hard-gap $\nu = 3/2$ FQH states hints that the two phases may be topologically distinct. One conceivable mechanism is anyon condensation, in which a bosonic anyon becomes gapless, condenses into the ground state, and thereby modifies the vacuum [42,43]. By way of illustration, the transition from a spin-singlet (331) state to a spin-polarized Pfaffian state has been described in such terms: a spin-flip bosonic anyon carrying charge $e/2$ and spin 1 condenses, closing the gap and freezing the spin degree of freedom [43]. Under a strong in-plane magnetic field, a spin-singlet (331) or anti-(331) state would be difficult to sustain; a spin-polarized $K = 8$ state might then emerge as an alternative [5]. Yet the $K = 8$ state appears incompatible with an anyon-condensation route, and it would not naturally produce a daughter state at $\nu = 1 + 6/13$, because its expected daughter states lie at $\nu = 8/15$ and $8/17$ [28].

That said, either a (331) or an anti-(331) state could still figure in the experimental phase diagram, because the in-plane magnetic field can in principle drive a $k$-space transition from one valley to two valleys in the GaAs/AlGaAs QW [39]. In such a picture, pseudospin - rather than spin-singlet pairing - would become the relevant degree of freedom, with pseudospin polarization progressively suppressed as the in-plane field grows. Provided inter-valley tunneling remains relatively strong, a one-component FQH state might well dominate.

It is also worth noting that the daughter state observed at $\nu = 19/13$ could point toward either an anti-Pfaffian or an anti-(331) parent state [28]. For an anti-Pfaffian state at $\nu = 1 + 1/2$, one expects daughter states at $\nu = 1 + 6/13 = 19/13$ and $\nu = 1 + 9/17 = 26/17$. For the anti-(331) state at $\nu = 2 - 1/2$ - the particle-hole conjugate of the (331) state - the predicted daughters would be $\nu = 2 - 7/13 = 19/13$ and $\nu = 2 - 9/19 = 29/19$. Thus the hard-gap FQH state might be interpreted as either anti-Pfaffian or anti-(331). The absence of the second anticipated daughter state may simply reflect its proximity to stronger neighboring phases or an energy gap below our experimental resolution.

A non-Abelian ground state remains a possibility to consider. Strong LL mixing between the $N = 0$ and $N = 1$ levels, induced by the in-plane magnetic field, would alter the effective pseudopotential at $\nu = 3/2$. At the same time, the rising Zeeman energy would further polarize the spin. Taken together, these effects might create conditions favorable for a paired CF state. The Coulomb anisotropy arising from broken rotational symmetry does not, however, appear strong enough in our experiment to produce pronounced transport anisotropy.

With these cautions in mind, one may sketch a possible interpretation of the two $\nu = 3/2$ FQH states. The hard-gap phase might, for instance, be ascribed to an anti-(331) state, stabilized when tunneling between the two Fermi-surface droplets becomes sufficiently weak, and accompanied by the daughter state at $\nu = 19/13$. As the in-plane magnetic field is reduced, a topological phase transition from this anti-(331) state to a paired CF state - such as the anti-Pfaffian - could conceivably occur. If so, this field-induced transition would differ from the (331)-to-Pfaffian transition reported at $\nu = 1/2$ in wide quantum wells, where the gap remains open throughout. Because the in-plane-field-induced anisotropy tends to weaken paired CF states, the resulting soft-gap phase would be only marginally stable. Unlike the $\nu = 5/2$ state in the SLL, where CF pairing is fragile and readily destroyed by in-plane-field anisotropy, the $\nu = 3/2$ state in the LLL may benefit from a larger CF effective mass and stronger LL mixing, rendering its pairing more resilient against anisotropic perturbation. Upon further reduction of the in-plane field, the system would then cross over from a developing soft-gap FQH state back to a CF liquid, signaling the underlying Fermi-surface transition.

Even within this tentative picture, the microscopic pairing mechanism of CFs remains to be understood, and the sharp increase of the $\nu = 3/2$ energy gap with field calls for a quantitative explanation. Moreover, the measured critical magnetic fields might not coincide with existing theoretical predictions. These lingering uncertainties remind us that the in-plane-field-driven phase transitions from a CF liquid to FQH states at $\nu = 3/2$ are still wide open to interpretation, and we hope the observations reported here will stimulate further theoretical ideas beyond the single scenario outlined above.

***Acknowledgements*** – Helpful discussions with Yang Liu, Junren Shi and Jun Hu are gratefully acknowledged. The work at Peking University was funded by the Innovation Program for Quantum Science and Technology - National Science and Technology Major Project (Grant No. 2021ZD0302600) and the National Key Research and Development Program of China (Grant No. 2024YFA1409002). The work at Princeton University was funded by the Gordon and Betty Moore Foundation through the EPiQS initiative Grant No. GBMF4420, by the National Science Foundation MRSEC Grant No. DMR-1420541.

## *References*

[1] R. L. Willett, et al., Observation of an even-denominator quantum number in the fractional quantum Hall effect, *Phys. Rev. Lett.* **59**, 17760-1779 (1987).

[2] G. Moore, & N. Read, Nonabelions in the fractional quantum Hall effect, *Nucl. Phys. B* **360**, 362-396 (1991).

[3] R. L. Willett, et al., Interference measurements of non-Abelian e/4 & Abelian e/2 quasiparticle braiding, *Phys. Rev. X* **13**, 011028 (2023).

[4] C. Nayak, S. H. Simon, A. Stern, M. Freedman, & S. D. Sarma, Non-Abelian anyons and topological quantum computation, *Rev. Mod. Phys.* **80**, 1083-1159 (2008).

[5] N. Read, & D. Green, Paired states of fermions in two dimensions with breaking of parity and time-reversal symmetries and the fractional quantum Hall effect, *Phys. Rev. B* **61**, 10267-10297 (2000).

[6] J. K. Jain, Composite Fermions, *Cambridge University Press* (2007).

[7] D. K. Ki, V. I. Fal'ko, D. A. Abanin, & A. F. Morpurgo, Observation of even denominator fractional quantum Hall effect in suspended bilayer graphene, *Nano Lett.* **14**, 2135-2139 (2014).

[8] J. Falson, et al., Even-denominator fractional quantum Hall physics in ZnO, *Nat. Phys.* **11**, 347-351 (2015).

[9] J. I. A. Li, et al., Even-denominator fractional quantum Hall states in bilayer graphene, *Science* **358**, 648-652 (2017).

[10] A. A. Zibrov, et al., Tunable interacting composite fermion phases in a half-filled bilayer-graphene Landau level, *Nature* **549**, 360-364 (2017).

[11] M. S. Hossain, et al., Unconventional anisotropic even-denominator fractional quantum Hall state in a system with mass anisotropy, *Phys. Rev. Lett.* **121**, 256601 (2018).

[12] Q. Shi, et al., Odd- and even-denominator fractional quantum Hall states in monolayer WSe2, *Nat. Nanotechnol.* **15**, 569-573 (2020).

[13] B. Dutta, et al., Distinguishing between non-Abelian topological orders in a quantum Hall system, *Science* **375**, 193-197 (2022).

[14] Y. W. Suen, L. W. Engel, M. B. Santos, M. Shayegan, & D. C. Tsui, Observation of a $\nu = 1/2$ fractional quantum Hall state in a double-layer electron system, *Phys. Rev. Lett.* **68**, 1379-1382 (1992).

[15] J. P. Eisenstein, G. S. Boebinger, L. N. Pfeiffer, K. W. West, & S. He, New fractional quantum Hall state in double-layer two-dimensional electron systems, *Phys. Rev. Lett.* **68**, 1383-1386 (1992).

[16] Y. W. Suen, H. C. Manoharan, X. Ying, M. B. Santos, & M. Shayegan, Origin of the $\nu = 1/2$ fractional quantum Hall state in wide single quantum wells, *Phys. Rev. Lett.* **72**, 3405 (1994).

[17] S. He, S. Das Sarma, & X. C. Xie, Quantized Hall effect and quantum phase transitions in coupled two-layer electron systems, *Phys. Rev. B* **47**, 4394-4412 (1993).

[18] B. I. Halperin, Theories for $\nu = 1/2$ in single- and double-layer systems, *Surf. Sci.* **305**, 1-7 (1994).

[19] M. Greiter, X. G. Wen, & F. Wilczek, Paired Hall states in double-layer electron systems, *Phys. Rev. B* **46**, 9586-9589 (1992).

[20] M. Greiter, X. G. Wen, & F. Wilczek, Paired Hall states, *Nucl. Phys. B* **374**, 567-614 (1992).

[21] M. R. Peterson, & S. D. Sarma, Quantum Hall phase diagram of half-filled bilayers in the lowest and the second orbital Landau levels: Abelian versus non-Abelian incompressible fractional quantum Hall states, *Phys. Rev. B* **81**, 165304 (2010).

[22] J. Shabani, et al., Phase diagrams for the stability of the $\nu = 1/2$ fractional quantum Hall effect in electron systems confined to symmetric, wide GaAs quantum wells, *Phys. Rev. B* **88**, 245413 (2013).

[23] N. Thiebaut, N. Regnault, & M. O. Goerbig, Fractional quantum Hall states versus Wigner crystals in wide quantum wells in the half-filled lowest and second Landau levels, *Phys. Rev. B* **92**, 245401 (2015).

[24] W. Zhu, Z. Liu, F. D. M. Haldane, & D. N. Sheng, Fractional quantum Hall bilayers at half filling: tunneling-driven non-Abelian phase, *Phys. Rev. B* **94**, 245147 (2016).

[25] A. Sharma, A. C. Balram, & J. K. Jain, Composite-fermion pairing at half-filled and quarter-filled lowest Landau level, *Phys. Rev. B* **109**, 035306 (2024).

[26] M. Levin, & B. I. Halperin, Collective states of non-Abelian quasiparticles in a magnetic field, *Phys. Rev. B* **79**, 205301 (2009).

[27] S. K. Singh, C. Wang, C. T. Tai, C. S. Calhoun, K. A. Villegas Rosales, P. T. Madathil, A. Gupta, K. W. Baldwin, L. N. Pfeiffer, & M. Shayegan, Topological phase transition between Jain states and daughter states of the ν = 1/2 fractional quantum Hall state, *Nat. Phys.* **20**, 1247 (2024).

[28] M. Yutushui, M. Hermanns, & D. F. Mross, Paired fermions in strong magnetic fields and daughters of even-denominator Hall plateaus, *Phys. Rev. B* **110**, 165402 (2024).

[29] E. Zheltonozhskii, A. Stern, & N. H. Lindner, Identifying the topological order of quantized half-filled Landau levels through their daughter states, *Phys. Rev. B* **110**, 245140 (2024).

[30] S. K. Singh, C. Wang, A. Gupta, K. W. Baldwin, L. N. Pfeiffer, & M. Shayegan, Fractional quantum Hall state at ν = 1/2 with energy gap up to 6K and possible transition from the one- to two-component state, *Phys. Rev. Lett.* **135**, 246603 (2025).

[31] S. Hasdemir, Y. Liu, H. Deng, M. Shayegan, L. N. Pfeiffer, K. W. West, K. W. Baldwin, & R. Winkler, ν = 1/2 fractional quantum Hall effect in tilted magnetic fields, *Phys. Rev. B* **91**, 045113 (2015).

[32] J. Shabani, Y. Liu, M. Shayegan, L. N. Pfeiffer, K. W. West, & K. W. Baldwin, Phase diagrams for the stability of the ν = 1/2 fractional quantum Hall effect in electron systems confined to symmetric, wide GaAs quantum wells, *Phys. Rev. B* **88**, 245413 (2013).

[33] R. R. Du, A. S. Yeh, H. L. Stormer, D. C. Tsui, L. N. Pfeiffer, and K. W. West, Fractional quantum Hall effect around ν = 3/2: composite fermions with a spin, *Phys. Rev. Lett.* **75**, 3926 (1995).

[34] T. S. Lay, T. Jungwirth, L. Smrčka, & M. Shayegan, One-component to two-component transition of the 2/3 fractional quantum Hall effect in a wide quantum well induced by an in-plane magnetic field, *Phys. Rev. B* **56**, R7092 (1997).

[35] J. Xia, V. Cvicek, J. P. Eisenstein, L. N. Pfeiffer, & K. W. West, Tilt-Induced Anisotropic to Isotropic Phase Transition at ν = 5/2, *Phys. Rev. Lett.* **105**, 176807 (2010).

[36] J. Xia, J. Eisenstein, L. Pfeiffer, & K. West, Evidence for a fractionally quantized Hall state with anisotropic longitudinal transport, *Nat. Phys.* **7**, 845 (2011).

[37] G. Liu, C. Zhang, D. C. Tsui, I. Knez, A. Levine, R. R. Du, L. N. Pfeiffer, & K. W. West, Enhancement of the ν = 5/2 Fractional Quantum Hall State in a Small In-Plane Magnetic Field, *Phys. Rev. Lett.* **108**, 196805 (2012).

[38] C. Wang, S. K. Singh, C. T. Tai, A. Gupta, L. N. Pfeiffer, K. W. Baldwin, & M. Shayegan, Stripe-Nematic Phase of Composite Fermions, *Phys. Rev. Lett.* **136**, 016501 (2026).

[39] L. Smrčka, & T. Jungwirth, The single-layer/bilayer transition of electron systems in AlGaAs/GaAs/AlGaAs quantum wells subject to in-plane magnetic fields, *J. Phys.: Cond. Matter* **7**, 3721 (1995).

[40] E. Bell, K. W. Baldwin, L. N. Pfeiffer, K. W. West, & M. A. Zudov, High-order two-component fractional quantum Hall states around filling factor ν = 1, arXiv:2512.04050.

[41] A. Kumar, G. A. Csáthy, M. J. Manfra, L. N. Pfeiffer, & K. W. West, Nonconventional Odd-Denominator Fractional Quantum Hall States in the Second Landau Level, *Phys. Rev. Lett.* **105**, 246808 (2010).

[42] F. A. Bais, & J. K. Slingerland, Condensate-induced transitions between topologically ordered phases, *Phys. Rev. B* **79**, 045316 (2009).

[43] M. Barkeshli, & X.-G. Wen, Anyon Condensation and Continuous Topological Phase Transitions in Non-Abelian Fractional Quantum Hall States, *Phys. Rev. Lett.* **105**, 216804 (2010).